\documentclass[preprint,12pt,nonatbib]{elsarticle}
\usepackage{soul}
\usepackage[T1]{fontenc}
\usepackage[utf8]{inputenc}
\usepackage[table]{xcolor}
\usepackage{graphicx}
\usepackage{subcaption}
\usepackage{amsmath,amssymb}
\usepackage{booktabs}
\usepackage{multirow}
\usepackage{svg}
\usepackage{textcomp}
\usepackage{adjustbox}
\usepackage{placeins}
\usepackage{tabularx}
\usepackage{hyperref}
\usepackage{cleveref}
\usepackage{url}
\usepackage{relsize}
\soulregister\cite7
\soulregister\autoref7

\newcommand{\hlmath}[1]{\mbox{$#1$}}
\soulregister\hlmath7

\DeclareUnicodeCharacter{2032}{\ensuremath{^{\prime}}}

\makeatletter
\let\c@author\relax
\makeatother

\usepackage[backend=biber,style=numeric,sorting=none,doi=false,url=false,isbn=false,eprint=false]{biblatex}

\renewbibmacro{in:}{}

\begin{document}

\begin{frontmatter}

\title{A high-dimensional neural network potential for finite-temperature phenomena in NiTi martensite}

\author[inst1]{Petr Jaroš}
\author[inst2]{Petr Sedlák\thanks{Corresponding author.}}
\ead{psedlak@it.cas.cz}
\author[inst2,inst3,inst4]{Petr Šesták}
\author[inst3,inst4]{Miroslav Černý}
\author[inst5,inst6]{Jörg Behler}
\author[inst2]{Hanuš Seiner}

\cortext[cor1]{Corresponding author.}

\address[inst1]{Department of Solid State Engineering, Faculty of Nuclear Sciences and Physical Engineering, Czech Technical University in Prague, 120 00 Prague, Czech Republic}
\address[inst2]{Institute of Thermomechanics, Czech Academy of Sciences, 182 00 Prague, Czech Republic}
\address[inst3]{Institute of Physical Engineering, Faculty of Mechanical Engineering, Brno University of Technology, 616 69 Brno, Czech Republic}
\address[inst4]{CEITEC BUT, Brno University of Technology, 612 00 Brno, Czech Republic}
\address[inst5]{Lehrstuhl für Theoretische Chemie II, Ruhr-Universität Bochum, 44780 Bochum,  Germany}
\address[inst6]{Research Center Chemical Sciences and Sustainability, Research Alliance Ruhr, 44780 Bochum, Germany}

\begin{abstract}
We present a high-dimensional neural network potential (HDNNP) for the martensitic phase of the NiTi shape-memory alloy trained to density functional theory (DFT) data. A central aspect of this work is the systematic validation of the potential with respect to the underlying DFT reference method for key properties governing structural evolution, including equilibrium crystal structures, elastic constants, generalized-stacking fault energies, and vibrational spectra. The HDNNP accurately describes the relative stability of the \texorpdfstring{B19$'$}{B19'} and B33 phases, including subtle energy differences on the order of meV/atom. The predicted stacking-fault energy landscape is strongly anisotropic and reveals a preferential shear pathway, providing atomistic insight into deformation and twinning mechanisms. Finite-temperature molecular dynamics simulations further enable the investigation of unconstrained structural evolution as a function of temperature. Overall, the developed HDNNP provides a robust basis for atomistic simulations of the complex structural and functional behavior of martensitic NiTi systems containing hundreds of thousands of atoms on nanosecond time scales.
\end{abstract}


\end{frontmatter}

\section{Introduction}

Nickel-titanium (NiTi) is the most widely used shape memory alloy (SMA). Despite the great attention that has been paid to this material for decades, several important properties such as ductility in polycrystalline materials, which are behind its success in technological applications, are not fully understood, which limits further systematic improvement and the search for new materials with similar functionality. The functional behavior of NiTi, i.e., the shape-memory effect and superelasticity derived from the reversible martensitic transformation of austenite (cubic B2 structure) to martensite (monoclinic  B19$'$  structure). The reversible functional response is, however, combined with exceptional ductility in NiTi, which is provided by additional plastic deformation mechanisms and is a key property for the use of this alloy in the applications. The plastic deformation mechanisms in NiTi have long been attributed mainly to the austenitic cubic phase \cite{chowdhury_significance_2016, ezaz_plastic_2013}, because this phase (unlike monoclinic martensite) has enough independent slip systems and therefore satisfies the generally accepted von Mises-Taylor criterion. This view has recently changed significantly, after systematic thermomechanical tests \cite{heller_beyond_2019}, so-called low-temperature shape setting experiments (LTSS) \cite{sittner_coupling_2018} and detailed high-resolution transmission electron microscopy (HRTEM) analysis of deformed martensitic microstructures \cite{molnarova_plastic_2023} clearly demonstrated that plastic deformation in martensite dominates over a wide temperature range. These new findings have motivated a deeper study of the fundamental properties of the B19$'$ martensitic structure. At the microstructural level, a mechanism for the unusual plasticity of monoclinic martensite has recently been proposed \cite{seiner_kwinking_2023}: based on a detailed theoretical analysis of experimentally observed plastically deformed martensitic microstructures, it has been proposed that plastic deformation in martensite proceeds as cooperative twinning and the formation and propagation of kink bands by a cooperative [100]$_M$(001)$_M$ slip mechanism. This unique combination of two known inelastic mechanisms has been termed “kwinking” and represents an unusual plastic deformation mechanism that gives martensite high ductility despite the involvement of only one slip system. However, the atomistic picture of this unusual plastic mechanism in the  B19$'$  structure is very incomplete.

Density functional theory (DFT), as a powerful tool for atomistic simulations in solids, has been used to calculate the properties of  B19$'$  in a number of studies
\cite{ko_temperature_2018,wagner_lattice_2008,guda_vishnu_phase_2010,hatcher_role_2009,huang_crystal_2003} and provided the first important insight: the  B19$'$  structure is not consistently predicted to be the energetic ground state. Instead, the ground state is an orthorhombic B33 structure, which does not satisfy the group-subgroup symmetry relationship between austenite and martensite. This brings a major complication in the further development of atomistic models. Due to computational demands, DFT calculations are mostly limited to small atomistic models, and the inclusion of finite temperature effects or deformation mechanisms in the structure is practically impossible. Therefore, no atomistic simulation of, e.g., the dislocation core in the  B19$'$  structure associated with the [100]$_M$(001)$_M$ easy slip has been realized so far. Recently, DFT calculations were able to demonstrate the thermal stabilization of the  B19$'$  structure at finite temperatures \cite{haskins_ab_2016,haskins_finite_2017}, providing a first estimate of the relative free energies of the B19' and B33 structures, which play an important role in the stability of B19' against slip as discussed in \cite{gao_origin_2017}. This was achieved either by means of a quasiharmonic approximation (involving sometimes also anharmonic corrections) and the calculation of the free energy evolution along the transformation path between the B33 and  B19$'$  structures \cite{ko_temperature_2018, wu_theoretical_2022}, or by applying ab-initio molecular dynamics \cite{haskins_finite_2017}. In the first case, the free energy was calculated based on the DFT energy and the vibrational spectra, however, due to computational complexity, the free energy evolution was mapped only along the transformation path defined by a single parameter (monoclinic angle of the  B19$'$  structure). In the latter, ab-initio molecular dynamics was used to simulate the time evolution of the structure at a finite temperature, but only for a small atomistic model (144-atom supercells—4×3×3 supercell of the 4 atom unit cells \cite{haskins_finite_2017}). More sophisticated DFT calculations are currently limited by their high computational cost.

In contrast, molecular dynamics (MD) simulations based on empirical potentials and force fields allow the investigation of large systems over longer timescales, but their predictive power strongly depends on the quality of the underlying interatomic potential. Bridging the gap between the accuracy of DFT and the efficiency of MD therefore represents a key challenge for atomistic modeling of NiTi. In this work, we address this challenge by constructing a machine learning potential \cite{behler_prespective, gabor_prespective, Noe_prespective, friederich_machine-learned_2021}, specifically a second-generation high-dimensional neural network potential (HDNNP) \cite{behler_generalized_2007, behler_4thgeneration} trained on DFT reference data, with the explicit aim of preserving DFT-level accuracy. Throughout the manuscript, we systematically assess the accuracy of the HDNNP by direct comparison with DFT calculations, focusing on properties that are critical for structural evolution with temperature. These include equilibrium crystal structures, elastic constants, generalized stacking fault energies, and vibrational properties. The HDNNP is focused on the martensitic B19′ phase and its relation to the B33 structure. At zero temperature, we demonstrate that the HDNNP reproduces the relative stability of these phases as well as the pronounced elastic anisotropy characteristic of B19$'$. The analysis of the generalized stacking-fault energy surface reveals a strong crystallographic anisotropy and identifies a single energetically favorable shear pathway along the [100]$_M$(001)$_M$ direction. As introduced in \cite{gao_origin_2017}, configurations closely related to the B33 structure emerge as local minima along this pathway, suggesting a shear-mediated structural connection between B19′ and B33. By allowing the system to relax freely at each temperature during MD simulations, we identify representative structures corresponding to free energy minima. A comparison between statically sheared configurations and structures obtained from molecular dynamics simulations shows that the latter are thermodynamically favored. This demonstrates the new possibility of an unconstrained search for stable structures at finite temperatures while maintaining the accuracy of DFT energy calculations.

\section{Methods}

\subsection{DFT setup}
DFT calculations were employed as a reference method for dataset generation. All total energies and atomic forces were computed using the Vienna \textit{Ab initio} Simulation Package (VASP)~\cite{kresse1996efficient,kresse1993ab,kresse1994ab,kresse1994norm,kresse1996efficiency}. The projector-augmented wave (PAW) method~\cite{kresse1999ultrasoft, blochl} was applied with a plane-wave energy cutoff of 650\,eV. Titanium \( p \)-electrons were explicitly treated as valence electrons. The exchange--correlation effects were described using the Perdew--Burke--Ernzerhof (PBE) functional within the generalized gradient approximation (GGA). To ensure isotropic sampling, the k-point mesh was chosen to maintain uniform spacing in all reciprocal lattice directions. For the B19$'$ primitive cell defined in \autoref{tab:Table_1}, a $24\times24\times24$ k-point grid was used. The Brillouin zone was sampled using a \(\Gamma\)-centered Monkhorst--Pack grid  with  the  Methfessel–Paxton broadening scheme with a smearing width of 0.05 eV. For larger supercells, the number of k-points was scaled accordingly to preserve approximately the same k-point density, thereby ensuring consistent sampling of reciprocal space. The electronic self-consistency criterion was set to an energy convergence threshold of \( 10^{-7} \,\text{eV} \). All these settings ensure the energy consistency below 1 meV/atom. The DFT results presented in this work were obtained using these computational settings.

\subsection{Dataset}\label{sec:dataset}

The dataset is a key component of any machine learning model. Unlike analytic potentials, which are derived from explicit physics-based equations, machine learning potentials learn the energies and forces from representative training structures, which numerical consistency achieved by constructing the forces as analytic energy derivatives. The dataset used in this work was constructed with a focus on two characteristic features of the martensitic phase of NiTi. The first is the discrepancy between the predicted and experimentally observed ground-state structures. Extensive DFT calculations \cite{huang_crystal_2003} suggest that the ground state of the martensitic phase of NiTi corresponds to an orthorhombic B33 structure, sometimes also referred to as a near-orthorhombic BCO structure. However, this phase has never been observed experimentally; instead, experiments consistently report the  B19$'$  structure \cite{kudoh_crystal_1985, prokoshkin_lattice_2004,buhrer_powder_1983}. The second motivation arises from experimental observations indicating a higher concentration of defects along a specific direction, namely in the (001) plane \cite{kudoh_crystal_1985}. The presence of a preferred slip system and strong anisotropy may play a crucial role in the plastic deformation process and has been further investigated in \cite{seiner_kwinking_2023}.

Motivated by these findings, we constructed a DFT-based dataset comprising 7300 structures, ranging from  4-atom primitive cells to  180-atom supercells. The structures were selected to capture possible transformation pathways between the B19$'$ and B33 phases, to represent strong elastic anisotropy, and to incorporate thermal fluctuations observed in molecular dynamics simulations in the temperature range of 0--700~K.

In the initial stage, the dataset consisted of structures generated by introducing random atomic displacements of  the relaxed atomic structure with magnitudes ranging from 0.1 to 0.2 Å along each coordinate using Atomsk \cite{hirel_atomsk_2015}. These perturbed structures served as the training set for the first HDNNP model. Using this preliminary model, we performed molecular dynamics (MD) simulations under various thermodynamic ensembles and conditions. During these simulations, the dataset was iteratively expanded with additional configurations selected according to two criteria: extrapolation parameters and an active learning strategy. Extrapolation refers to situations in which local atomic environments extend beyond the boundaries defined by the atom-centered symmetry function values serving as descriptors of the atomic environments \cite{behler_atom-centered_2011}  of configurations already included in the dataset. Incorporating such configurations allowed the model to systematically broaden the domain of its training data.

Once sufficient configurational coverage had been achieved, the dataset was further refined through an active learning process designed to minimize prediction errors. Specifically, we adopted the committee model methodology as our active learning strategy \cite{schran_committee_2020, tokita_how_2023}. The committee comprised seven HDNNPs  (referred to as members), summarized in \autoref{tab:Table_2}, each trained on the same dataset but differing in network architecture—namely in the number of neurons—and in the specific partitioning of the dataset into training and test subsets. Large discrepancies among the predictions of the committee members indicated that a given local atomic environment was poorly represented in the current dataset and should therefore be included in the retraining set. Because atomic forces serve as more sensitive indicators of model accuracy than total energies, we analyzed the MD trajectories and extracted the configurations exhibiting the largest force discrepancies, while preserving periodic boundary conditions, and added them to the dataset. This procedure was repeated iteratively until the standard deviation of the committee disagreement for each force component was less than $0.03~\mathrm{eV}/$\AA.

\begin{table}[htbp]
    \centering
    \begin{tabular}{llcccc}
    \toprule
       structure & method & \textit{ a} [\AA]& \textit{b} [\AA] & \textit{c} [\AA] & $\beta$ $[^{\circ}]$ \\
    \midrule
        B33 & HDNNP    & 2.937 & 4.017 & 4.915  &  107.0 \\
        B33 & DFT  & 2.933 &  4.011 & 4.918 & 107.0   \\
        B33  & DFT \cite{huang_crystal_2003}&  2.940  & 3.997  & 4.936 & 107.0 \\
        B33  & 2NN-MEAM \cite{KoMEAM}&  3.149  & 3.760  & 4.984 & 108.4 \\
        B19$'$ & Exp. \cite{kudoh_crystal_1985} & 2.898 & 4.108 & 4.646 & 97.78  \\
        B19$'$& HDNNP    & 2.918 & 4.063 &  4.680  & 97.78   \\
        B19$'$& DFT & 2.926 & 4.060 &  4.635  & 97.78   \\
        B19$'$ & DFT  \cite{wagner_lattice_2008} & 2.941& 4.035 & 4.685 & 97.78 \\
        B19$'$  & 2NN-MEAM \cite{KoMEAM}&  2.878  & 4.129  & 4.659 & 99.4 \\

    \bottomrule
    \end{tabular}\caption{Comparison of structure parameters predicted by the presented HDNNP, DFT calculations and  data presented in the literature. Structure parameters for B33 were calculated from minimization of all degrees of freedom while the structure parameters for  B19$'$  were determined through minimization with fixed monoclinic angle of the unit cell. The structures are visualized in \autoref{fig:figure_3}.}
    \label{tab:Table_1}
\end{table}

\subsection{High-dimensional neural network potential}
The reference dataset consists of potential energies, \(E_{\mathrm{ref}}\), and atomic forces, \(F_{\mathrm{ref}}\), obtained from DFT calculations. In the HDNNP formalism \cite{behler_generalized_2007,behler_4thgeneration}, the potential energy of the training system, \(E_{\mathrm{ref}}\), is calculated as a sum of environment-dependent atomic energies. Individual atomic forces are determined as analytic derivatives of the system potential energy with respect to the atomic coordinates.

The construction of the HDNNP involves three essential components: a reference dataset, a descriptor of local atomic environments, and a fitting procedure used to optimize the neural network parameters. The local atomic environments are described using atom-centered symmetry functions (ACSFs)~\cite{behler_atom-centered_2011}, which characterize the neighborhood of each atom within a finite cutoff radius, which was set here to \(R_c = 6.36~\text{\AA}\). This cutoff radius ensures an accurate representation of the local atomic environment. The local environments are encoded using a combination of radial and angular ACSFs, which ensure invariance of the potential with respect to translation, rotation, and permutation of identical atoms.

Firstly, a series of fully connected neural network models with two hidden layers, with identical architecture for each element, were fitted. The models summarised in \autoref{tab:Table_2} were trained using the same dataset and fitting procedure, but differed slightly in the number of neurons in the hidden layers and in the composition of the training and test subsets. The model with 34 and 31 neurons in the first and second hidden layer, respectively, exhibited the lowest force RMSE and was therefore selected as the final model discussed in this work. The remaining models were employed as members of the committee model used in the active learning procedure.

\begin{figure}[htbp]
    \centering
    \includegraphics[width=1\textwidth]{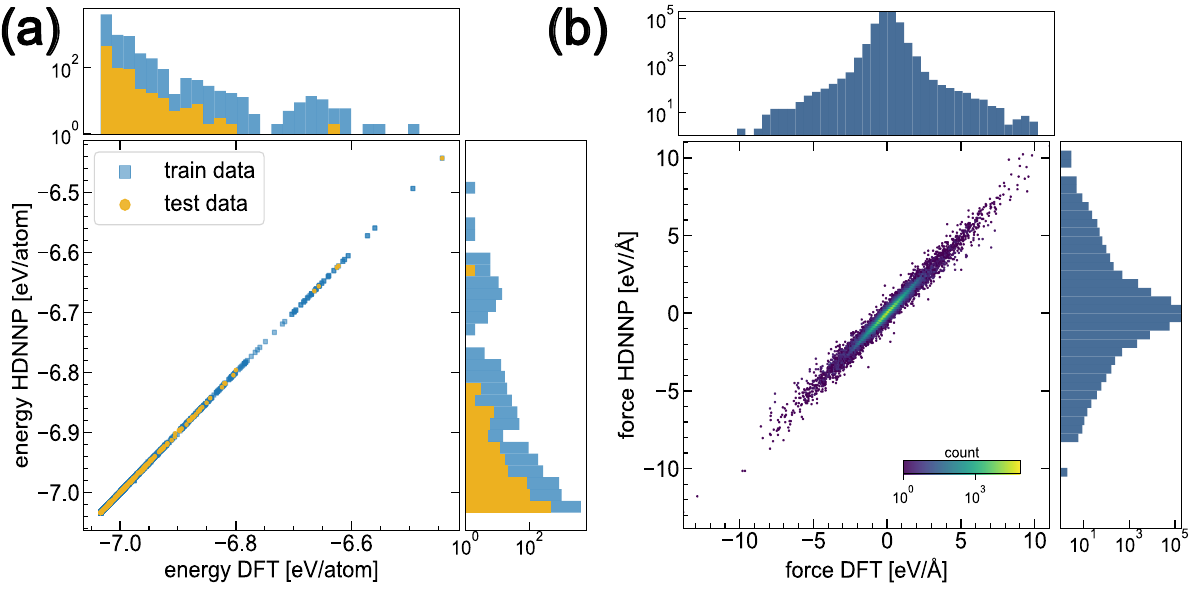}
    \caption{Correlation graphs between (a) energies and (b) forces predicted by the HDNNP and reference data obtained from DFT. The marginal histograms denote the statistical distribution of the data, corresponding to the number of configurations within each energy or force interval for the reference DFT and HDNNP-predicted values.}
    \label{fig:figure_1}
\end{figure}

For each chemical element, the ACSFs consisted of 72 descriptors. Ten percent of the dataset was reserved for testing, while the remaining configurations were used for training. The training was performed for 30 epochs using the Kalman filter optimizer~\cite{runner_kalman}; in each epoch, 10\% of the force components were selected randomly for fitting procedure. The resulting model achieved root mean square error (RMSE) values of 0.34 and 0.43\,meV/atom for the training and test energies, respectively, and 0.074 and 0.078\,eV/\AA\ for the training and test force components, respectively. The model with all ACSFs parameters and specification of all fitting parameters is available at \cite{model_zenodo}. The quality of the fit is shown in \autoref{fig:figure_1}, demonstrating good agreement between the HDNNP predictions and the DFT reference data.

The neural network potential was constructed using RuNNer version 1.3~\cite{behler_constructing_2015,behler_first_2017} and interfaced with molecular dynamics simulations in LAMMPS~\cite{LAMMPS} through the n2p2 package~\cite{singraber_library-based_2019,singraber_parallel_2019}. Further details on ACSFs and the HDNNP fitting procedure are provided in \ref{appendix_1}.
\begin{table}[htbp]
\centering

\small
\setlength{\tabcolsep}{4pt}

\begin{adjustbox}{max width=\textwidth}
\begin{tabular}{c c c c c}
\toprule
\multirow{2}{*}{\textbf{Hidden layers}} &
\multicolumn{2}{c}{Energy [meV/atom]} &
\multicolumn{2}{c}{Forces [eV/\AA]} \\
\cmidrule(lr){2-3} \cmidrule(lr){4-5}
& Train RMSE & Test RMSE & Train RMSE & Test RMSE \\ \midrule
33--32 & 0.364 & 0.884 & 0.0760 & 0.0651 \\
31--31 & 0.410 & 0.443 & 0.0792 & 0.0641 \\
33--32 & 0.342 & 0.490 & 0.0743 & 0.0772 \\
31--29 & 0.346 & 0.457 & 0.0745 & 0.0788 \\
33--29 & 0.324 & 0.930 & 0.0750 & 0.0863 \\
29--29 & 0.368 & 0.481 & 0.0768 & 0.0747 \\
30--29 & 0.349 & 0.415 & 0.0743 & 0.0724 \\
34--31$^*$ & 0.340 & 0.430 & 0.0740 & 0.0780 \\
\bottomrule \end{tabular}
\end{adjustbox}
\caption{RMSE values of the fitted neural network models with two hidden layers. The column Hidden layers specifies the model topology, where the notation $n_1$--$n_2$ denotes the number of neurons in the first and second hidden layer, respectively. The model marked by an asterisk is the presented model discussed in this paper. The remaining models were used as members of the committee model in the active learning approach. All models were trained using the same fitting procedure, while differing in the composition of the training and test subsets and in the number of neurons.}
\label{tab:Table_2} \end{table}

\section{Validation of HDNNP}

In this section, we examine whether the HDNNP model reproduces the key structural characteristics obtained from DFT that are relevant to the temperature-driven evolution of the martensitic phase. Rather than focusing solely on quantitative agreement in energies and forces, we aim to verify that the HDNNP captures the same physical trends and relative phase stabilities as the reference DFT calculations. To this end, HDNNP predictions are compared with our own DFT results and with available literature data, with particular emphasis on equilibrium structures as well as the elastic and vibrational properties of the B19$'$ and B33 phases. This validation establishes a physically consistent reference state for the finite-temperature simulations discussed in the following sections. Validation of the HDNNP with respect to DFT calculations demonstrates that the HDNNP methodology is applied consistently, which can then be considered as a method that enables the upscaling of DFT calculations and can still be considered as an ab-initio method without any empirical inputs. The comparison of properties predicted by interatomic potentials with experiments is not performed at this point, but will be addressed in the Discussion section.

\subsection{Equilibrium structures}
\begin{figure}[htbp]
    \centering
\includegraphics[scale=0.6]{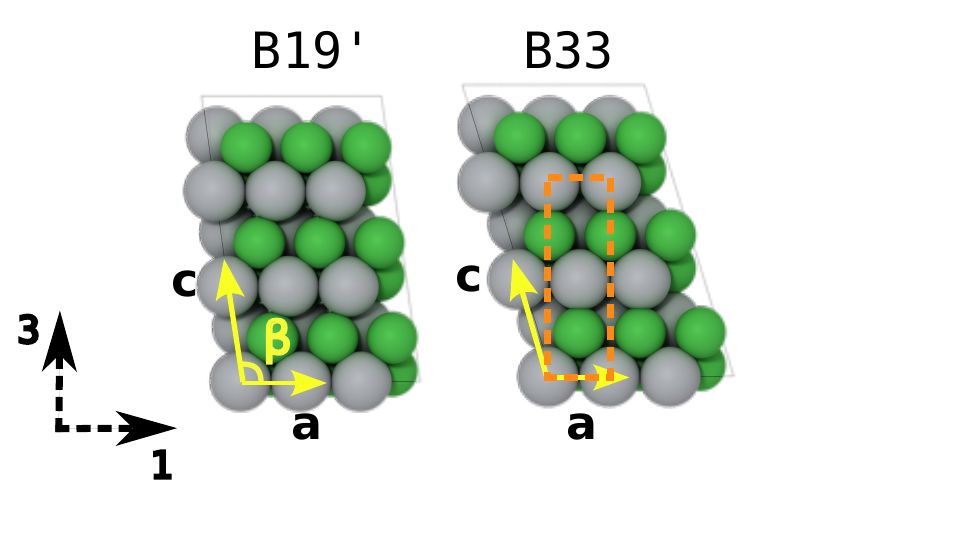}
    \caption{Visualization of the B19$'$ and B33 structures projected onto the (010) plane, highlighting the monoclinic angle $\beta$. The B33 phase exhibits orthorhombic symmetry, indicated by the orange dashed rectangle. Titanium atoms are shown in gray, while nickel atoms are represented in green. The visualization was generated using OVITO \cite{ovito}.}
    \label{fig:figure_3}
\end{figure}

We first consider the equilibrium structures of the B19$'$ and B33 phases which are visualized in \autoref{fig:figure_3}. Structural parameters obtained from HDNNP relaxations and from DFT calculations are compared with reported literature values to assess whether the HDNNP reproduces the DFT relative energies and characteristic lattice parameters of the two phases. In the DFT calculations, the B33 structure is obtained through full energy relaxation starting from the experimentally reported B19$'$ unit cell summarised in \autoref{tab:Table_1}. All atomic positions, lattice parameters, and cell angles are allowed to relax, allowing the system to evolve toward its energetic minimum at 0~K. The same procedure is followed using the HDNNP in LAMMPS. The structural relaxation is dominated by an increase in the monoclinic angle \(\beta\) from \(97.78^\circ\), characteristic of the experimentally observed B19$'$ phase, to approximately \(107.0^\circ\), corresponding to the B33 structure. To determine the relaxed structural parameters of the B19$'$ phase itself, an additional energy minimization is carried out using both HDNNP and DFT, while keeping the lattice angles constrained. The resulting lattice parameters for both phases are summarized in \autoref{tab:Table_1}. In both DFT and HDNNP calculations, this result indicates that the two approaches identify the same ground-state structure and corresponding energy minimum. At the same time, it is seen in \autoref{tab:Table_1} that the 2NN-MEAM potentials, while predicting correct structures in terms of lattice symmetry, deviate considerably from the DFT predictions in the lattice parameters. This brings an additional motivation for translating the DFT precision into molecular dynamics potentials through machine learning, so that the molecular dynamics simulations would approximate more closely the B33 and B19$'$ lattices.

\begin{figure}[htbp]
    \centering
    \includegraphics[width=\linewidth]{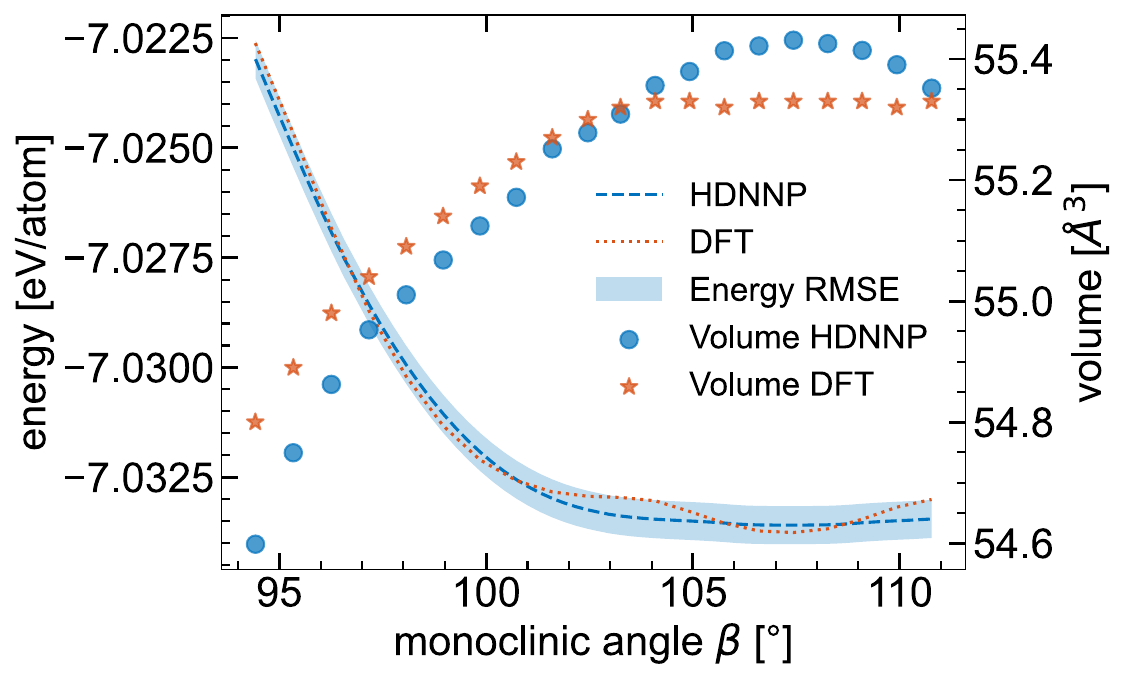}
    \caption{The energies of the B19$'$ and B33 phases were evaluated as a function of the monoclinic angle $\beta$. For each prescribed value of $\beta$, the structure, including the lattice parameters other than the prescribed angle, was relaxed using either HDNNP or DFT, and the potential energy was calculated. The accuracy of the HDNNP was quantified using the RMSE of the fit.}
    \label{fig:figure_2}
\end{figure}

Having established agreement for the fully relaxed structures, we next examine whether the HDNNP provides a consistent description of intermediate configurations connecting B19$'$ and B33. To this end, the path connecting both structures is sampled by systematically varying the monoclinic angle $\beta$. This variation is introduced through a pure stress shear applied along the crystallographic direction [100]$_M$(001)$_M$ (see \autoref{fig:figure_3}), which directly modifies the monoclinic angle and therefore provides a connection between B19$'$ and B33 structures. Such a transition was already discussed in \cite{mizuno_compositional_2015, guda_vishnu_phase_2010}. For each intermediate cell shape, an energy relaxation was first performed using DFT. The resulting relaxed structure was then used as the initial configuration for a subsequent HDNNP minimization, carried out under the same conditions. The resulting energy profiles are shown in \autoref{fig:figure_2}. The results of HDNNP and DFT predictions are comparable with those presented  in \cite{mizuno_compositional_2015,guda_vishnu_phase_2010, huang_crystal_2003} agreeing with transition pathway between  B19$'$  and B33 ground state. The transition between the B19$'$ and B33 structures is accompanied by a change in the unit-cell volume, with the B33 phase exhibiting a larger volume, as shown in \autoref{fig:figure_2}. This result may appear counterintuitive, given that the B33 structure represents the DFT-derived ground state at 0~K.

\subsection{Elastic tensor}
\begin{figure}[htbp]
    \centering
    \includegraphics[width=1\textwidth]{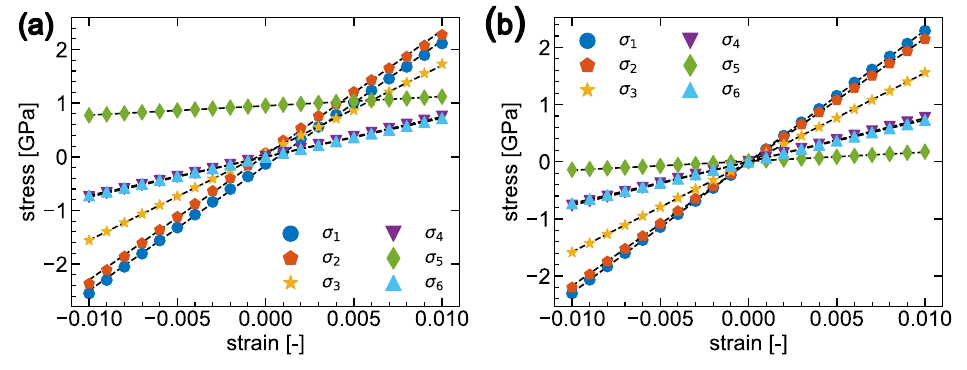}
    \caption{Stress--strain relations for (a) B19$'$ and (b) B33. The points represent the stresses calculated using HDNNP, while the dashed lines correspond to linear fits. The individual stress components, \(\sigma_i\), are plotted as a function of the corresponding strains, \(\epsilon_i\). In total, 36 stress--strain curves were obtained; for clarity, only those corresponding to the main diagonal components of the elastic tensor are shown. }
    \label{fig:figure_4}
\end{figure}

Having established that the HDNNP reproduces the equilibrium structures and relative stability of the B19$'$ and B33 phases, we now turn to their elastic response at 0~K. Elastic properties provide a sensitive test of the potential, as they probe the curvature of the energy landscape around the equilibrium configuration and are directly linked to the pronounced anisotropy of the martensitic phase. The relaxed equilibrium structures of B19$'$ and B33 used as starting configurations are summarized in \autoref{tab:Table_1}. To determine the full elastic tensor, six independent strains always with only one nonzero strain component $\varepsilon_1,\ldots,\varepsilon_6$ (in Voigt notation), were applied sequentially. The corresponding six stress components, $\sigma_1,\ldots,\sigma_6$, were then evaluated, using LAMMPS for the HDNNP and VASP for the DFT calculations. For calculations using the HDNNP and DFT, strains in the range of $-1\%$ to $+1\%$ were applied in order to remain within the linear elastic regime. After each deformation step, atomic positions were relaxed at fixed cell shape. The calculation resulted in 36 stress-strain relations that were fitted with linear functions, the slopes of these fits yield the elastic constants. The calculated stress--strain curves representing the relation for the diagonal parameters of the elastic tensor are shown in  \autoref{fig:figure_4} for the phases B19$'$ and B33.  The resulting elastic constants are summarized in \autoref{tab:Table_3}. The HDNNP reproduces the strong elastic anisotropy of B19$'$, reflected in the low $C_{55}$ elastic constant, in agreement with DFT results and with values reported in the literature \cite{wagner_lattice_2008, haskins_ab_2016}. The non-zero shift in $\sigma_5$ at zero strain in \autoref{fig:figure_4} shows that B19' is not in the energy minimum and a non-zero shear is necessary to stabilize it as previously discussed in \cite{wagner_lattice_2008}.

\begin{table}[htbp]
\centering

\adjustbox{max width=\textwidth}{
\begin{tabular}{l c c c c c c c c}
\toprule
Elastic constant  &  B19$'$  & B19$'$   & B19$'$ &  B19$'$  & B19$'$  &    B33 &  B33  & B33\\
{ }{[GPa]} & HDNNP   & DFT  & DFT  \cite{wagner_lattice_2008} & DFT   \cite{haskins_finite_2017} & 2NN-MEAM \cite{KoMEAM} & HDNNP   & DFT  & 2NN-MEAM \cite{KoMEAM}\\
\midrule
$C_{11}$ & 144 & 191 & 195 & 177 & 188 &164 & 189 & 156\\
$C_{12}$ & 127 & 126 & 128 & 132 &88 &125 & 125 & 121\\
$C_{13}$ & 119 & 102 & 96  & 114 & 105 &119 & 98 & 77 \\
$C_{15}$ & 13  & 10  & 13  & 5  & -23  &0  & 1 &  0   \\
$C_{22}$ & 216 & 244 & 241 & 240 & 276 &219 & 233 & 190 \\
$C_{23}$ & 135 & 122 & 126 & 128 & 42  &122 & 128 &  93\\
$C_{25}$ & -4  & -6 & -9  & -9  & 47 &1.3  &  0 &  0  \\
$C_{33}$ & 252 & 238 & 235 & 234 & 262 &253 & 240 &  222   \\
$C_{35}$ & 19  & 14   & 20 & 1  & -35 &3 & 2 &  0  \\
$C_{44}$ & 69  & 77  & 76  &77  & 87 &73  & 93 & 82  \\
$C_{46}$ & 2   & 6 & -4  & 5  & -4  &1 & 0 &  0 \\
$C_{55}$ & 14  & 22  & 21  & 14  & 107  & 1  & 7 & 71  \\
$C_{66}$ & 73  & 78  & 77  & 80  & 114  &75  & 84 &  125  \\
\bottomrule
\end{tabular}
}
\caption{Comparison of the calculated elastic constants, $C_{ij}$, with literature values at 0~K, using the coordinate system and lattice structure defined in \autoref{fig:figure_3}. Remarkably, the DFT and HDNNP both confirm strong elastic instability of the B19’ with respect to [100](001) shears, as represented by the elastic constant $C_{55}$; for the 2NN-MEAM \cite{KoMEAM} potential, in contrast, $C_{55}$ is comparable to $C_{44}$ and $C_{66}$, indicating a much more isotropic shear behavior of the model lattice.}
\label{tab:Table_3}
\end{table}

\subsection{Generalized stacking-fault energy}
\begin{figure}[h!]
    \centering
    \includegraphics[width=1.0\textwidth]{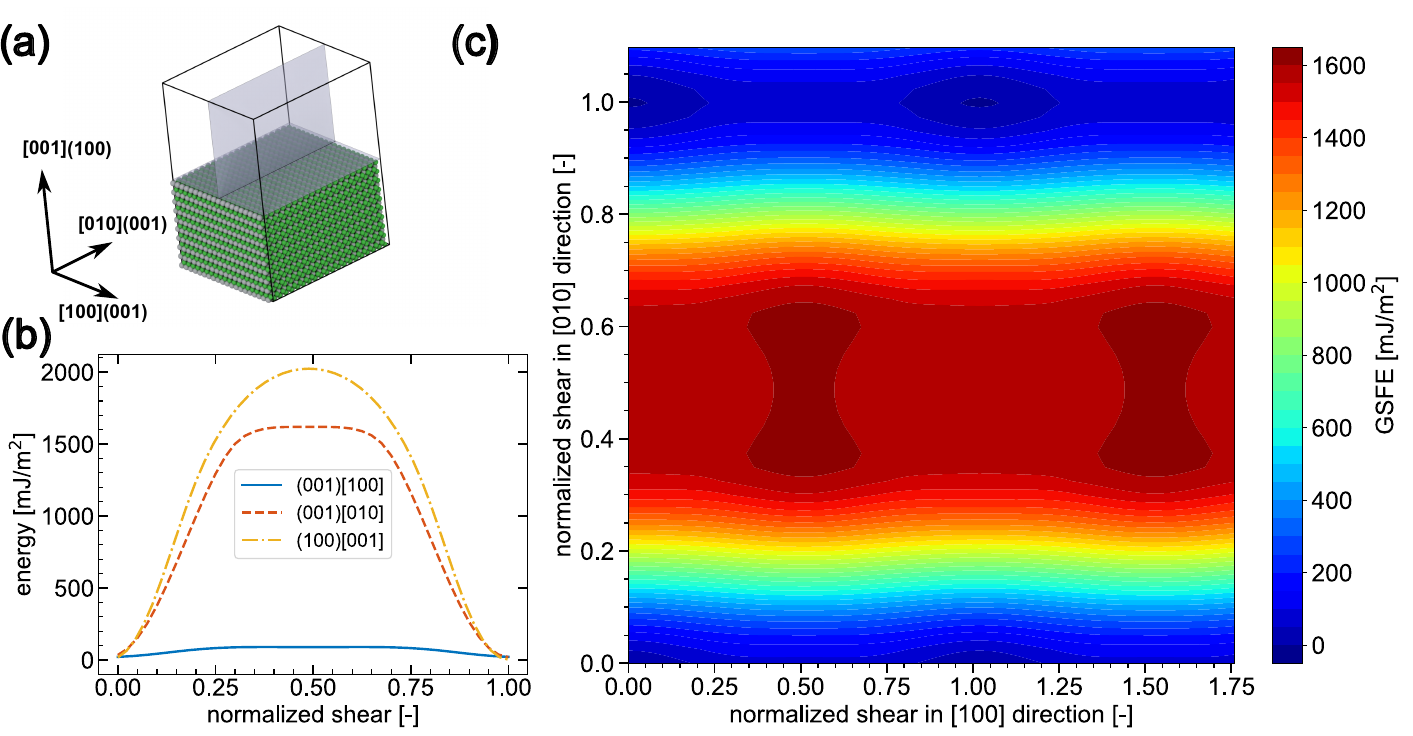}
    \caption{(a) The visualization of the shear planes with respect to the used crystallographic directions. (b) The comparison of the GSFE along the three main crystallographic directions of B19$'$. (c) GSFE in the (001) plane shows the energy barriers across the entire plane. The results indicate that only one energetically favorable shear pathway exists, along the  [100] direction.
 }
    \label{fig:figure_5}
\end{figure}

The pronounced elastic anisotropy of the B19$'$ martensite, discussed in the previous section, indicates that shear-related deformation mechanisms may play a central role in its mechanical response. To further investigate this behavior, we examine the generalized stacking-fault energy (GSFE) \cite{vitek_intrinsic_1968} using the HDNNP. The GSFE provides direct insight into the energetic barriers associated with shear and  therefore provides insight into the crystallographically allowed slip systems. The GSFE was calculated using a $10\times6\times20$ supercell with 4800 atoms, which was divided into two blocks of equal size. One block was incrementally displaced relative to the other by $0.1$~\AA\ steps along the chosen shear direction. During this procedure, the displaced block was kept rigid, while the atomic positions in the second block were allowed to relax only in the direction perpendicular to the shear plane. The GSFE of the B19$'$ structure was evaluated along three characteristic shear directions: [100] and [001] within the (001) plane, and [001] within the (100) plane. The resulting energy landscapes, shown in \autoref{fig:figure_5}, exhibit a pronounced directional dependence, reflecting the intrinsic anisotropy of the martensitic lattice. The systematic calculation of all possible shear paths within the (001) plane is shown in \autoref{fig:figure_5}, demonstrating that, within the (001) plane, only a single energetically favorable slip pathway exists, namely along the [100] direction. All other shear directions exhibit significantly higher energy barriers. These results highlight a strong anisotropy of the GSFE and indicate that plastic deformation and transformation-related shear in B19$'$ are highly constrained crystallographically. The identification of a unique low-energy slip path along the [100]$_M$(001)$_M$ direction is consistent with previous DFT studies \cite{ezaz_higher_2012} and  may be coupled with the emergence of a local minimum similar to B33 discussed in \cite{seiner_kwinking_2023, gao_symmetry_2019}.

\subsection{Phonon analysis at 0~K}
Once it has been shown that the HDNNP    reproduces the equilibrium structures, elastic response, and shear-related anisotropy of the B19$'$ and B33 phases, we now turn to vibrational properties. A consistent description of lattice dynamics is essential if DFT-level accuracy is to be transferred to molecular dynamics simulations.

\begin{figure}[h!]
    \centering
    \includegraphics[width=0.8\linewidth]{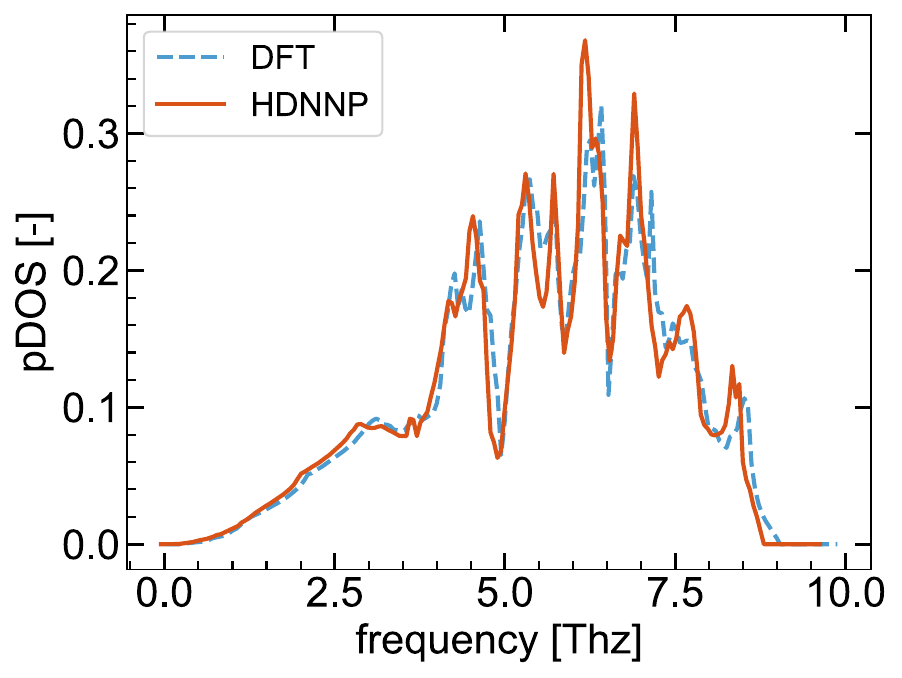}
    \caption{Comparison of the phonon density of states (pDOS) of the B33 phase obtained from structure relaxation using the presented HDNNP and DFT calculations. Both pDOS were computed on a $3\times3\times3$ supercell using the dynamical matrix approach.}
    \label{fig:figure_6}
\end{figure}

\begin{figure}[h!]
    \centering
    \includegraphics[width=0.8\linewidth]{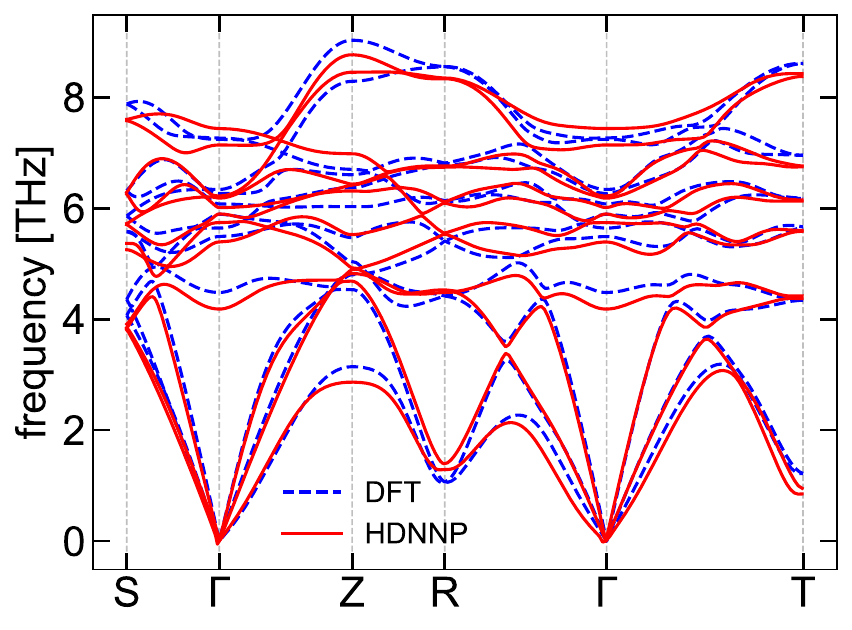}
    \caption{Comparison of the phonon dispersions of the B33 phase obtained from structure relaxation using the HDNNP and DFT calculations. Both dispersions were computed on a $3\times3\times3$ supercell using the dynamical matrix approach.}
    \label{fig:figure_7}
\end{figure}

The vibrational properties were evaluated in terms of the phonon density of states (pDOS) and phonon dispersion relations at 0~K. Both quantities were computed using the Phonopy package \cite{phonopy-phono3py-JPCM, phonopy-phono3py-JPSJ}, employing the dynamical matrix approach. In this framework, the force constants are obtained from finite atomic displacements, and the dynamical matrix is defined as
\begin{equation}\label{equation_1}
D_{\alpha\beta}(ij) = \frac{1}{\sqrt{m_i m_j}}
\frac{\partial^2 E}{\partial r_{i\alpha}\,\partial r_{j\beta}},
\end{equation}
where $E$ is the total potential energy, $m_i$ and $m_j$ are atomic masses, and $r_{i\alpha}$ and $r_{j\beta}$ denote Cartesian displacements of atoms $i$ and $j$. The phonon frequencies are then obtained by solving the eigenvalue problem associated with this matrix. Within this framework, we computed the pDOS of the relaxed B33 structure using the HDNNP and DFT. In both cases, a $3\times3\times3$ supercell containing 108 atoms was employed. The resulting pDOS spectra are compared in \autoref{fig:figure_6}. The close correspondence between the two curves indicates that the HDNNP reproduces the vibrational spectrum of the B33 ground state obtained by DFT accurately. While the pDOS captures an integral measure of the vibrational spectrum, it does not resolve individual phonon branches. To further assess whether the HDNNP    accurately reproduces the fundamental vibrational modes, we computed the phonon dispersion relations along the high-symmetry directions of the Brillouin zone corresponding to the B33 symmetry. The comparison between HDNNP and DFT dispersions is shown in \autoref{fig:figure_7}, demonstrating that the HDNNP reproduces not only the overall vibrational density but also the detailed phonon branch structure of the ground state.

\section{Comparison of HDNNP and semi-empirical potential}

In this section, we further illustrate the motivation for developing an HDNNP interatomic potential specifically tailored to martensitic NiTi by comparing its predictions with those of a widely used semi-empirical potential based on the second-nearest-neighbor modified embedded atom method (2NN-MEAM). To the best of the authors' knowledge, existing interatomic potentials for NiTi have been developed primarily to describe the austenite--martensite transformation, associated phase stability, and interfacial properties, whereas the energetics of deformation pathways within the martensitic phase have received considerably less attention. Consequently, their predictive capability for configurations associated with such deformation modes may be limited.

For this comparison, we selected the widely used 2NN-MEAM potential developed by Ko et al.~\cite{KoMEAM}. This potential reproduces both the B19$'$ and B33 structures and, consistently with DFT, predicts the orthorhombic B33 structure as the martensitic ground state. However, its equilibrium lattice parameters differ from the DFT predictions by several percent, and the predicted monoclinic angle, $\beta = 108.4^\circ$, is slightly larger than the corresponding DFT values summarized in \autoref{tab:Table_1}.

\begin{figure}[h!]
    \centering
    \includegraphics[width=1.0\linewidth]{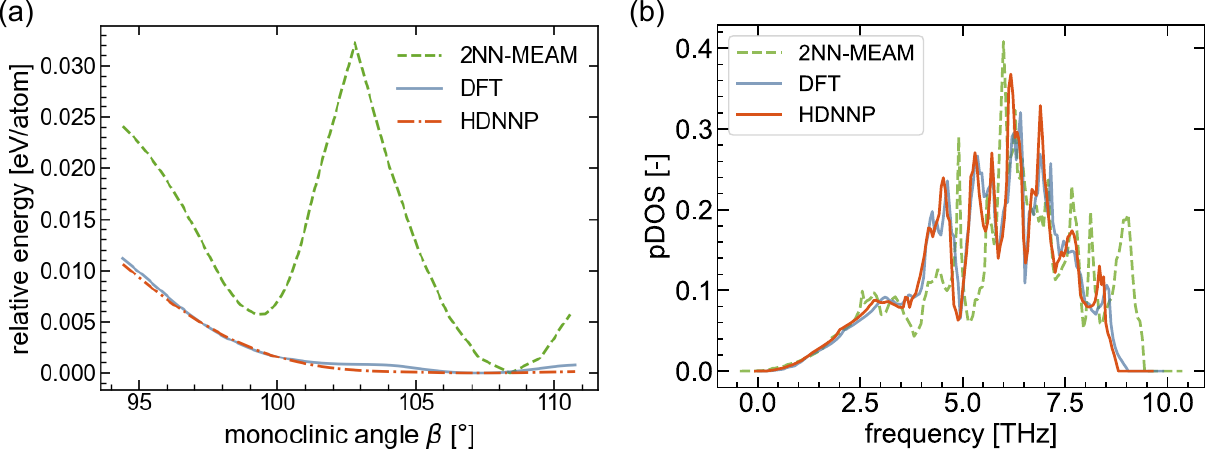}
    \caption{(a) Energy of structures along the B19$'$--B33 deformation path as a function of the monoclinic angle $\beta$, as predicted by the present HDNNP and the 2NN-MEAM interatomic potential of Ko et al. \cite{KoMEAM}. For each potential, the energies are given relative to the corresponding predicted ground state. (b) Comparison of the pDOS of the relaxed B33 phase obtained using the HDNNP and 2NN-MEAM potentials. The pDOS was calculated for a $3\times3\times3$ supercell using the dynamical-matrix approach.}
    \label{fig:figure_8}
\end{figure}

\autoref{fig:figure_8}a shows the energy landscape of intermediate structures connecting the B19$'$ and B33 configurations through variation of the monoclinic angle $\beta$. While the DFT calculations and the HDNNP simulations result in smooth evolutions, overcoming zero or very small energy barriers, the energy landscape predicted by the 2NN-MEAM potential is more complex. Within the 2NN-MEAM description, the B19$'$ structure corresponds to a local energy minimum separated by an energy barrier from the global minimum associated with the B33 structure. The structural parameters corresponding to these minima are listed in \autoref{tab:Table_1}. The energy branches associated with the B19$'$-like and B33-like configurations near the respective minima are shown in  \autoref{fig:figure_8}a. The two branches intersect at approximately $\beta = 103^\circ$, where configurations with nearly identical energies but different lattice parameters are obtained.

The absence of low-energy pathway connecting the B19$'$ and B33 structures and, consequently, of the related shear instability in the B19′ phase, is also  reflected in the relatively large value of the $C_{55}$ elastic constant from the 2NN-MEAM calculation seen in \autoref{tab:Table_3}. This behavior differs substantially from the much softer shear response obtained from DFT calculations \cite{wagner_lattice_2008, haskins_finite_2017} and reproduced by the present HDNNP. Differences are also observed in the phonon density of states of the B33 structure (\autoref{fig:figure_8}b). However, because the B19$'$ structure is locally stabilized in the 2NN-MEAM description, the B33 vibrational spectrum plays a less direct role in finite-temperature simulations of B19$'$ stability than in the DFT-based description.

Taken together, these observations indicate that the 2NN-MEAM potential provides a less accurate description of the energetics and structural response of configurations intermediate between B19$'$ and B33. This may limit its suitability for simulations of plastic-deformation mechanisms involving the corresponding martensitic shear modes, such as those proposed in recent works~\cite{gao_origin_2017,seiner_kwinking_2023} and discussed in the Introduction. This limitation should, however, be considered in the context of the original scope and parameterization of the 2NN-MEAM potential, which was not specifically designed to resolve this deformation pathway. The comparison therefore highlights the importance of explicitly including such martensitic deformation modes in the reference dataset when constructing an interatomic potential intended for simulations of the finite-temperature behavior and deformation mechanisms of NiTi martensite.

\section{Properties of the HDNNP at finite temperature}
\label{Relation between monoclinic angle and temperature}

\begin{figure}[htbp]
    \centering
    \includegraphics[width=1.0\linewidth]{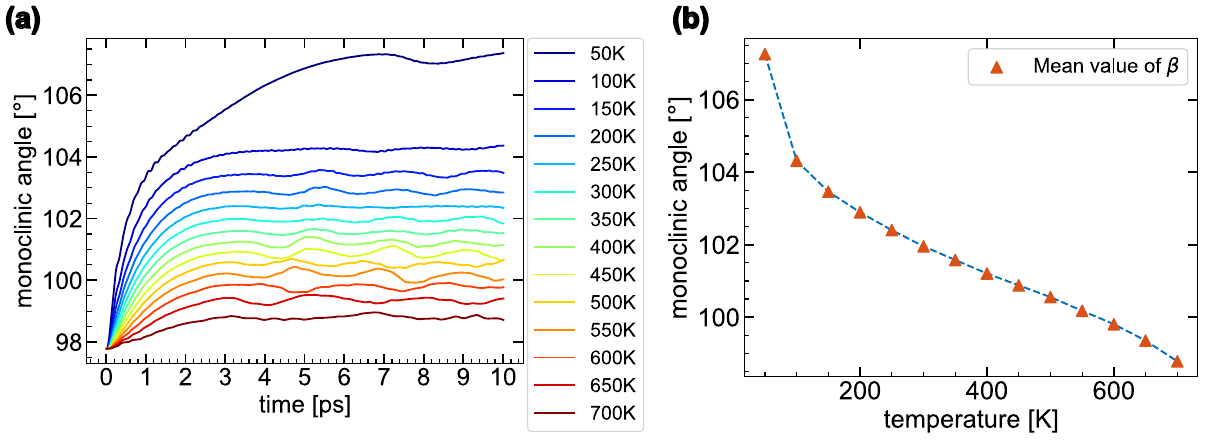}

    \caption{Evolution of the monoclinic angle, as obtained from HDNNP-driven molecular dynamics simulations performed in the \textit{NpT} ensemble, with the lattice parameters, atomic positions and unit-cell angles treated as degrees of freedom.
    (a) Time evolution of the monoclinic angle during the MD simulation.
    (b) Mean value of the monoclinic angle $\beta$ after structural relaxation at each temperature.}
    \label{fig:figure_9}
\end{figure}

\begin{figure}[htbp]
    \centering
    \includegraphics[width=1\textwidth]{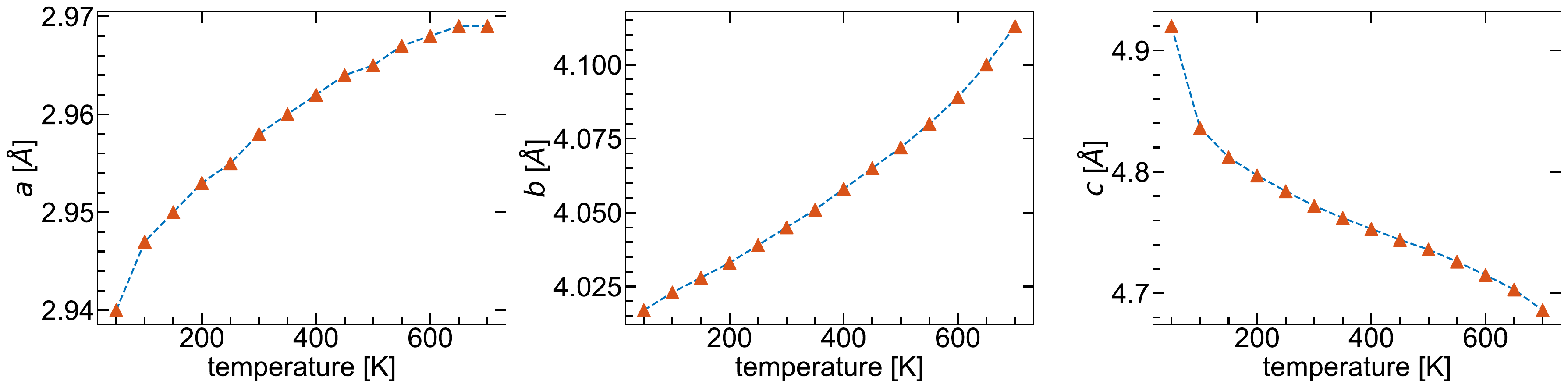}
    \caption{The temperature evolution of the mean primitive-cell parameters obtained from the NpT MD simulations. }
    \label{fig:figure_10}
\end{figure}

\begin{figure}[h!]
    \centering

    \includegraphics[width=1\textwidth]{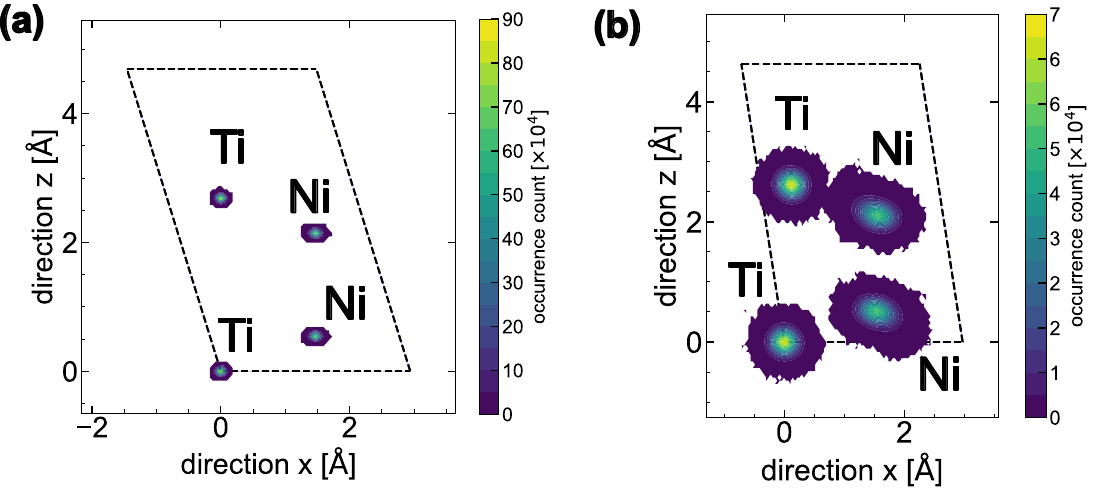}

    \caption{The primitive cells at (a) 50~K and (b) 700~K. The color bar indicates the spatial distribution of atoms around their mean positions.}
    \label{fig:figure_11}
\end{figure}

\begin{figure}[h!]
    \centering
    \includegraphics[width=1.0\textwidth]{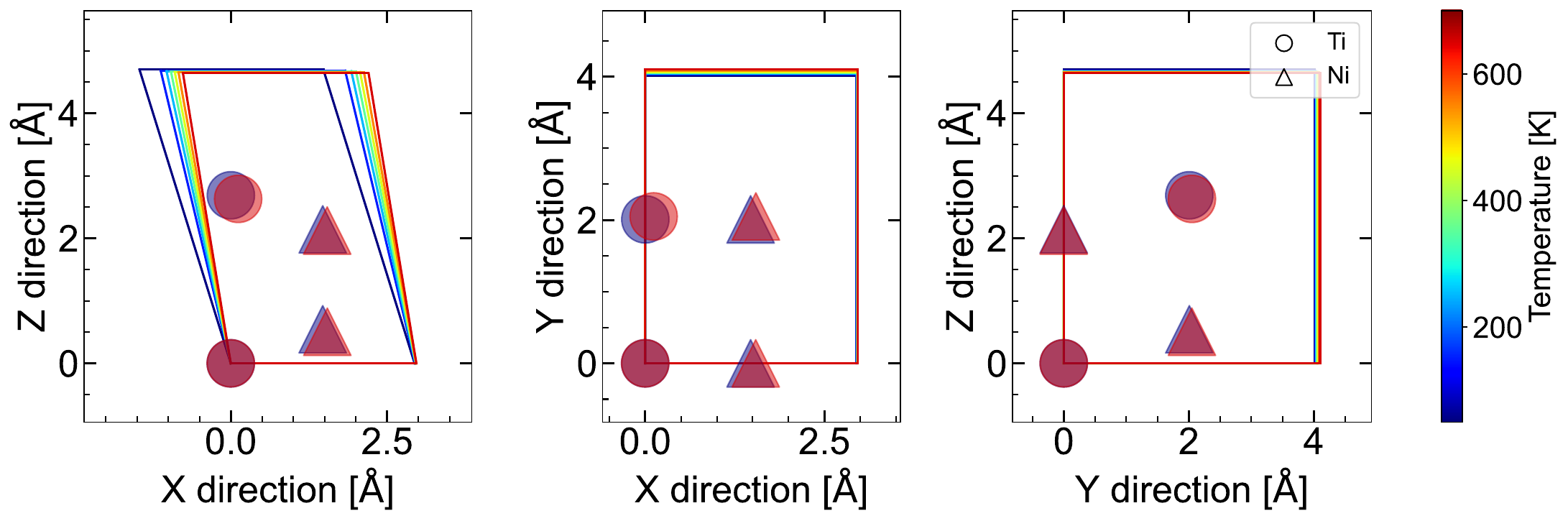}
    \caption{Temperature-induced evolution of lattice parameters and atomic positions. The figure shows projections of atomic coordinates onto the XZ, XY, and YZ planes. Marker colors distinguish structures between 50~K and 700~K according to the color bar.}
    \label{fig:figure_12}
\end{figure}

Up to this point, the results obtained at $T = 0$~K show good agreement with DFT calculations. The key advantage of the HDNNP framework, however, lies in its ability to extend this level of accuracy to system sizes and timescales beyond DFT approaches. To investigate the influence of temperature on the stability and structural evolution of the B19$'$ phase, we performed MD simulations in LAMMPS over a broad range of temperatures. During these simulations, the evolution of lattice parameters and cell angles was monitored continuously. The simulation cell contained 32,000 atoms and was constructed by replicating the experimental primitive cell parameters summarized in \autoref{tab:Table_1}, 20 times along each Cartesian direction. All simulations were carried out in the \textit{NpT} ensemble using a Nosé-Hoover barostat under a zero stress field. The run lasted 15 ps with time step of 1~fs. All simulation box parameters, including cell lengths and angles, were allowed to relax freely. Structural relaxation was typically achieved within approximately 8~ps, after which the system remained stable. Structural changes were examined at temperatures ranging from 50~K to 700~K. The average lattice parameters and cell angles were obtained from the mean value of supercell's parameters. Because the instantaneous supercell dimensions allowed the translation vectors to be identified at each time step, the average atomic positions and their associated thermal fluctuations could be determined at each temperature. \autoref{tab:Table_4} summarizes the resulting lattice parameters, monoclinic angles, and atomic positions of the reconstructed primitive cell. Reported values correspond to mean quantities and their standard deviations obtained from the MD trajectories. The temperature dependence of the monoclinic angle is shown in \autoref{fig:figure_9}, while \autoref{fig:figure_10} and \autoref{fig:figure_11} illustrate the variation of the lattice parameters and atomic positions. Representative atomic configurations at selected temperatures are shown in \autoref{fig:figure_12}. The structural change is mainly driven by a shear-like mechanism  of the atomic blocks of two Ni and two Ti atomic layers in the [100]$_M$(001)$_M$ direction.

\begin{table}[p]
\centering

{\scriptsize
\setlength{\tabcolsep}{2pt}
\renewcommand{\arraystretch}{0.95}

\begin{adjustbox}{max totalsize={\textwidth}{0.90\textheight},center}
\begin{tabular}{ccccccc}
\toprule
Temperature [K] & Label & Cell & Element & x [\AA] & y [\AA] & z [\AA] \\
\midrule

         \multirow{4}{*}{50}  & \textit{a} &2.940 $\pm$ 0.000 & Ti & 0.000 $\pm$ 0.041  &  0.000 $\pm$ 0.036 &  0.000 $\pm$ 0.032  \\
                               & \textit{b} &4.017 $\pm$ 0.001 & Ti &  0.002 $\pm$ 0.042 & 2.009 $\pm$ 0.035  &  2.678 $\pm$ 0.032  \\
                               & \textit{c} &4.920 $\pm$ 0.012 & Ni &  1.472 $\pm$ 0.051 &  0.000 $\pm$ 0.029 & 2.143 $\pm$ 0.032  \\
                               & $\beta$ &  107.257 $\pm$ 0.057 & Ni & 1.141 $\pm$ 0.051  & 2.009 $\pm$ 0.029  &  0.545 $\pm$ 0.032 \\
                                \hline
         \multirow{4}{*}{100} & \textit{a} &2.947 $\pm$ 0.001 & Ti & 0.000 $\pm$ 0.053  &  0.000 $\pm$ 0.051 & 0.000 $\pm$ 0.046  \\
                               & \textit{b} &4.023 $\pm$ 0.001 & Ti & 0.055 $\pm$ 0.053  & 2.011 $\pm$ 0.051  &  2.675 $\pm$ 0.046 \\
                               & \textit{c} &4.836 $\pm$ 0.005 & Ni & 1.508 $\pm$ 0.069  &  0.000 $\pm$ 0.042  & 2.136 $\pm$ 0.047  \\
                               & $\beta$ & 104.314 $\pm$ 0.069 & Ni & 1.494 $\pm$ 0.069  & 2.011 $\pm$ 0.042  & 0.539 $\pm$ 0.047  \\
                               \hline
         \multirow{4}{*}{150} & \textit{a} &2.95 $\pm$ 0.001 & Ti & 0.000 $\pm$ 0.061  &  0.000 $\pm$ 0.036 &  0.000 $\pm$ 0.058 \\
                               & \textit{b} &4.028 $\pm$ 0.001 & Ti & 0.070 $\pm$ 0.062  &  2.014 $\pm$ 0.063  &  2.670 $\pm$ 0.058 \\
                               & \textit{c} &4.812 $\pm$ 0.007 & Ni & 1.518 $\pm$ 0.082  &  0.000 $\pm$ 0.052 &  2.133 $\pm$ 0.058 \\
                               & $\beta$ & 103.462 $\pm$ 0.074  & Ni &  1.502 $\pm$ 0.082 &  2.014 $\pm$ 0.052 &  0.537 $\pm$ 0.058\\
                               \hline
         \multirow{4}{*}{200} & \textit{a} &2.953 $\pm$ 0.001  & Ti & 0.000 $\pm$ 0.071  &  0.000 $\pm$ 0.073  & 0.000 $\pm$ 0.067  \\
                               & \textit{b} &4.033 $\pm$ 0.002  & Ti & 0.079 $\pm$ 0.071  & 2.017 $\pm$ 0.074  & 2.667 $\pm$ 0.067   \\
                               & \textit{c} &4.797 $\pm$ 0.007 & Ni & 1.524 $\pm$ 0.096  & 0.000 $\pm$ 0.06  &  2.131 $\pm$ 0.068  \\
                               & $\beta$ & 102.894 $\pm$ 0.077 & Ni & 1.508 $\pm$ 0.095  & 2.017 $\pm$ 0.06  &  0.535 $\pm$0.068  \\
                               \hline
         \multirow{4}{*}{250} & \textit{a} &2.955 $\pm$ 0.001 & Ti & 0.000 $\pm$ 0.077  & 0.000 $\pm$ 0.083  & 0.000 $\pm$ 0.076  \\
                               & \textit{b} &4.039 $\pm$ 0.002 & Ti & 0.087 $\pm$ 0.077  & 2.020 $\pm$ 0.083  & 2.662 $\pm$ 0.076  \\
                               & \textit{c} &4.784 $\pm$ 0.004 & Ni & 1.529 $\pm$ 0.106  & 0.000 $\pm$ 0.068  & 2.130 $\pm$ 0.076  \\
                               & $\beta$ & 102.41 $\pm$ 0.08 & Ni & 1.513 $\pm$ 0.105  & 2.020 $\pm$ 0.068  & 0.533 $\pm$ 0.077  \\
                               \hline
         \multirow{4}{*}{300} & \textit{a} &2.958 $\pm$ 0.001 & Ti & 0.000 $\pm$ 0.088  & 0.000 $\pm$ 0.092  &  0.000 $\pm$ 0.084  \\
                               & \textit{b} &4.045 $\pm$ 0.002 & Ti & 0.093 $\pm$ 0.088  &  2.023 $\pm$ 0.092 & 2.659 $\pm$ 0.084  \\
                               & \textit{c} &4.772 $\pm$ 0.005 & Ni & 1.533 $\pm$ 0.119  & 0.000 $\pm$ 0.076  & 2.129 $\pm$ 0.085  \\
                               & $\beta$ & 101.955 $\pm$ 0.083 & Ni & 1.518 $\pm$ 0.119  &  2.023 $\pm$ 0.075  &  0.530 $\pm$ 0.085  \\
                               \hline
         \multirow{4}{*}{350} & \textit{a} &2.96 $\pm$ 0.001 & Ti & 0.000 $\pm$ 0.092  & 0.000 $\pm$ 0.100  & 0.000 $\pm$ 0.089  \\
                               & \textit{b} &4.051 $\pm$ 0.002 & Ti & 0.098 $\pm$ 0.092  & 2.026 $\pm$ 0.100  & 2.656 $\pm$ 0.090  \\
                               & \textit{c} &4.762 $\pm$ 0.007 & Ni & 1.536 $\pm$ 0.127  &  0.000 $\pm$ 0.082 & 2.128 $\pm$ 0.091  \\
                               & $\beta$ & 101.58 $\pm$ 0.086 & Ni & 1.522 $\pm$ 0.127  & 2.026 $\pm$ 0.082  & 0.529 $\pm$0.091  \\
                               \hline
         \multirow{4}{*}{400} & \textit{a} &2.962 $\pm$ 0.001 & Ti &  0.000 $\pm$ 0.098 &  0.000 $\pm$ 0.108 & 0.000 $\pm$ 0.097  \\
                               & \textit{b} &4.058 $\pm$ 0.002 & Ti & 0.103 $\pm$ 0.098  & 2.029 $\pm$ 0.108  & 2.653 $\pm$ 0.097  \\
                               & \textit{c} &4.753 $\pm$ 0.005 & Ni & 1.538 $\pm$ 0.136  & 0.000 $\pm$ 0.089  & 2.127 $\pm$ 0.099  \\
                               & $\beta$ & 101.206 $\pm$ 0.089 & Ni & 1.527 $\pm$ 0.136  & 2.029 $\pm$ 0.089  & 0.526 $\pm$ 0.099  \\
                               \hline
         \multirow{4}{*}{450} & \textit{a} &2.964 $\pm$ 0.001 & Ti & 0.000 $\pm$ 0.105  & 0.000 $\pm$ 0.116  & 0.000 $\pm$ 0.103  \\
                               & \textit{b} &4.065 $\pm$ 0.002 & Ti & 0.107 $\pm$ 0.105  &  2.032 $\pm$ 0.116  & 2.650 $\pm$ 0.103  \\
                               & \textit{c} &4.744 $\pm$ 0.01 & Ni & 1.540 $\pm$ 0.146    &  0.000 $\pm$ 0.095 &  2.127 $\pm$ 0.106 \\
                               & $\beta$ & 100.88 $\pm$ 0.093 & Ni &  1.530 $\pm$ 0.146  & 2.032 $\pm$ 0.095  &  0.524 $\pm$ 0.106 \\
        \hline
         \multirow{4}{*}{ 500} & \textit{a} & 2.965 $\pm$ 0.002& Ti & 0.000 $\pm$ 0.113  &  0.000 $\pm$ 0.122  &  0.000 $\pm$ 0.110\\
                               & \textit{b} & 4.072 $\pm$ 0.002 & Ti & 0.110 $\pm$ 0.113  & 2.036 $\pm$ 0.123  &  2.647 $\pm$ 0.110  \\
                               & \textit{c} & 4.736 $\pm$ 0.008& Ni &  1.541 $\pm$ 0.155 &   0.000 $\pm$ 0.102 &  2.126 $\pm$ 0.113 \\
                               & $\beta$ & 100.555  $\pm$ 0.095  & Ni & 1.534 $\pm$ 0.156  &  2.036 $\pm$ 0.102 & 0.521 $\pm$ 0.113  \\
        \hline
         \multirow{4}{*}{550} & \textit{a} &2.967 $\pm$ 0.001 & Ti & 0.000 $\pm$ 0.117  &  0.000 $\pm$ 0.129 &  0.000 $\pm$ 0.116 \\
                               & \textit{b} &4.08 $\pm$ 0.002 & Ti &  0.113 $\pm$ 0.117 & 2.040 $\pm$ 0.129  &  2.643 $\pm$ 0.116 \\
                               & \textit{c} & 4.726 $\pm$ 0.009 & Ni & 1.542 $\pm$ 0.163  & 0.000 $\pm$ 0.107  &  2.126 $\pm$ 0.119 \\
                               & $\beta$ & 100.178 $\pm$ 0.099 &   Ni &  1.538 $\pm$ 0.163  &  2.040 $\pm$ 0.107  & 0.517 $\pm$ 0.119  \\
                               \hline
         \multirow{4}{*}{600} & \textit{a} &2.968 $\pm$ 0.001 & Ti & 0.000 $\pm$ 0.126  &  0.000 $\pm$ 0.135 &  0.000 $\pm$ 0.123 \\
                               & \textit{b} &4.089 $\pm$ 0.003 & Ti & 0.116 $\pm$ 0.126  & 2.044 $\pm$ 0.135  &  2.637 $\pm$ 0.123 \\
                               & \textit{c} &4.715 $\pm$ 0.006 & Ni & 1.543 $\pm$ 0.173  &  0.000 $\pm$ 0.113 & 2.125 $\pm$ 0.128  \\
                               & $\beta$ & 99.808 $\pm$ 0.093 & Ni &  1.541 $\pm$ 0.172 &  2.045 $\pm$ 0.113  & 0.512 $\pm$ 0.127  \\
                               \hline
         \multirow{4}{*}{650} & \textit{a} &2.969 $\pm$ 0.001 & Ti & 0.000 $\pm$ 0.132  & 0.000 $\pm$ 0.142  & 0.000 $\pm$ 0.129  \\
                               & \textit{b} &4.100 $\pm$ 0.003 & Ti & 0.118 $\pm$ 0.132  &  2.050 $\pm$ 0.143 &  2.630 $\pm$ 0.129 \\
                               & \textit{c} &4.703 $\pm$ 0.008 & Ni &  1.542 $\pm$ 0.181 & 0.000 $\pm$ 0.12  & 2.124 $\pm$ 0.134  \\
                               & $\beta$ & 99.357 $\pm$ 0.097 & Ni & 1.544 $\pm$ 0.181  &  2.050 $\pm$ 0.12  &  0.506 $\pm$ 0.134 \\
                               \hline
         \multirow{4}{*}{700} & \textit{a} &2.969 $\pm$ 0.002  & Ti & 0.000 $\pm$ 0.139  & 0.000 $\pm$ 0.149  & 0.000 $\pm$ 0.135  \\
                               & \textit{b} &4.113 $\pm$ 0.003 & Ti & 0.119 $\pm$ 0.139  & 2.056 $\pm$ 0.148  &  2.619 $\pm$ 0.135  \\
                               & \textit{c} &4.686 $\pm$ 0.013  & Ni & 1.541 $\pm$ 0.189  & 0.000 $\pm$ 0.126  & 2.123 $\pm$ 0.141  \\
                               & $\beta$ & 98.784 $\pm$ 0.116  & Ni & 1.547 $\pm$ 0.189  &   2.057 $\pm$ 0.126 &  0.496 $\pm$ 0.141 \\
\bottomrule
\end{tabular}
\end{adjustbox}
}
\caption{The structural evolution of lattice parameters and atomic positions in the Cartesian coordinate system at different temperatures. The parameters \textit{a,b} and \textit{c} are in \AA, the monoclinic angle $\beta$ is in  degrees.  }
\label{tab:Table_4}
\end{table}

\section{Discussion}

In the previous sections, we demonstrated that the developed HDNNP provides a consistent description of both ground-state properties and finite-temperature behavior when applied in molecular dynamics simulations. At 0~K, the predicted structural, elastic, and vibrational properties are in close agreement with DFT calculations. At finite temperatures, the MD simulations reveal a gradual structural evolution from configurations close to B33 at lower temperatures toward configurations approaching B19$'$ with increasing temperature, driven by a shear-like mechanism. For each temperature, a representative mean structure is identified and summarized in \autoref{tab:Table_4}.

\begin{figure}[h!]
    \centering
    \includegraphics[width=1\textwidth]{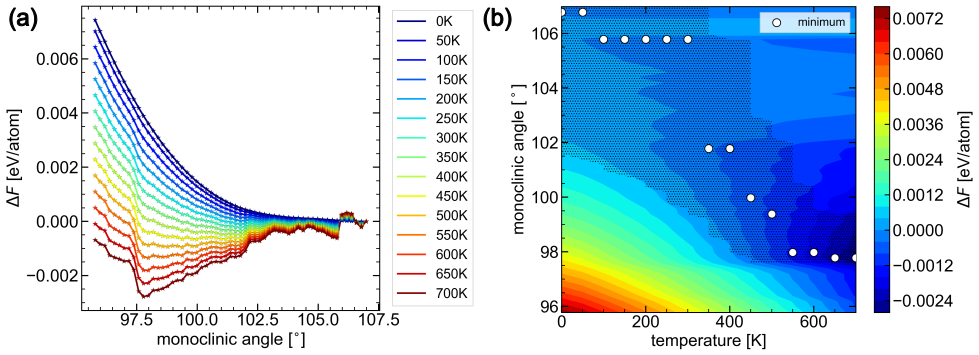}
    \caption{(a) Free energy difference as a function of monoclinic angle and temperature. The free energy difference is calculated relative to the B33 (corresponding to $107^{\circ}$) structure at each temperature. (b) Free energy difference as a function of temperature and monoclinic angle. The hatched region represents all accessible minima within the prediction error, while the white points indicate the  free-energy minima at each temperature obtained from (a).}
    \label{fig:figure_13}
\end{figure}

To quantify this structural evolution, we evaluated the free energy of two distinct sets of structures: (i) structures generated by a gradual static change of the monoclinic angle as was done in \autoref{fig:figure_2} (hereafter referred to as shifted structures), and (ii) mean structures obtained from MD simulations. For the shifted structures, the free energy was determined as a function of temperature and monoclinic angle. In contrast, for the MD-derived mean structures, the free energy was evaluated only as a function of temperature, since each temperature uniquely defines the corresponding monoclinic angle. In both cases, the free energy was calculated within the harmonic approximation~\cite{Dove1993},
\begin{equation}\label{equation_4}
F = \phi
+ \frac{1}{2}\hbar \int g(\omega)\,\omega \, d\omega
+ k_B T \int g(\omega)\,
\ln\!\left( 1 - \exp\!\left( -\frac{\hbar \omega}{k_B T} \right) \right)
\, d\omega ,
\end{equation}
where $\phi$ denotes the potential energy, $\omega$ is the phonon frequency, and $g(\omega)$ is the phonon density of states. Although the results are reported for temperatures up to 700~K, anharmonic contributions were not explicitly included. Previous work \cite{wu_theoretical_2022} has shown, using thermodynamic integration for the B19$'$ phase at 600~K, that the anharmonic contribution to the free energy amounts to approximately $-0.16$~meV/atom. This value lies within the prediction error of the present HDNNP model and is therefore not expected to affect the conclusions of this study. For both the shifted structures and the mean structures obtained from MD simulations, the phonon density of states was calculated using the dynamical matrix approach with $6\times6\times6$ supercells. \autoref{fig:figure_13} shows the evolution of the free-energy minima for the shifted structures over the temperature range from 0 to 700~K. The data are presented as free-energy differences relative to the DFT ground-state (B33) structure at each temperature. With increasing temperature, the position of the free-energy minimum gradually shifts toward the experimentally observed monoclinic angle of $97.7^\circ$, reaching this value at temperatures of approximately 450~K and above. For clarity, the data are presented without error bars. However, uncertainties play an important role, as the free-energy landscape exhibits shallow minima. Here, we consider the prediction error of the potential energy to be the main source of uncertainty. Therefore, for the free energy difference, we estimate the error as $\sqrt{2}\sigma_{\mathrm{RMSE}}$, which is equal to 0.61 meV/atom.

To visualize all free-energy minima that fall within the prediction uncertainty, we constructed the free-energy surface as a function of temperature and monoclinic angle. The resulting energy landscape for the shifted structures is shown in \autoref{fig:figure_13}. The hatched region indicates all accessible minima within the estimated prediction error, while the white symbols denote the global free-energy minima at each temperature.

As can be seen from \autoref{fig:figure_13}, the range of accessible monoclinic angles narrows with increasing temperature. The experimentally observed monoclinic angle becomes accessible from approximately 450~K up to 700~K, in agreement with a previous DFT-based study that followed the same connecting path \cite{ko_temperature_2018}. Nevertheless, the free-energy landscape in \autoref{fig:figure_13} shows an increasing change in curvature around the experimentally observed monoclinic angle of 97.7$^\circ$, starting at approximately 250~K. This suggests that the 97.7$^\circ$ configuration may play a role in the transformation by forming an energy barrier after an otherwise relatively flat region of the free-energy landscape.

\begin{figure}[h!]
    \centering
    \includegraphics[width=1\textwidth]{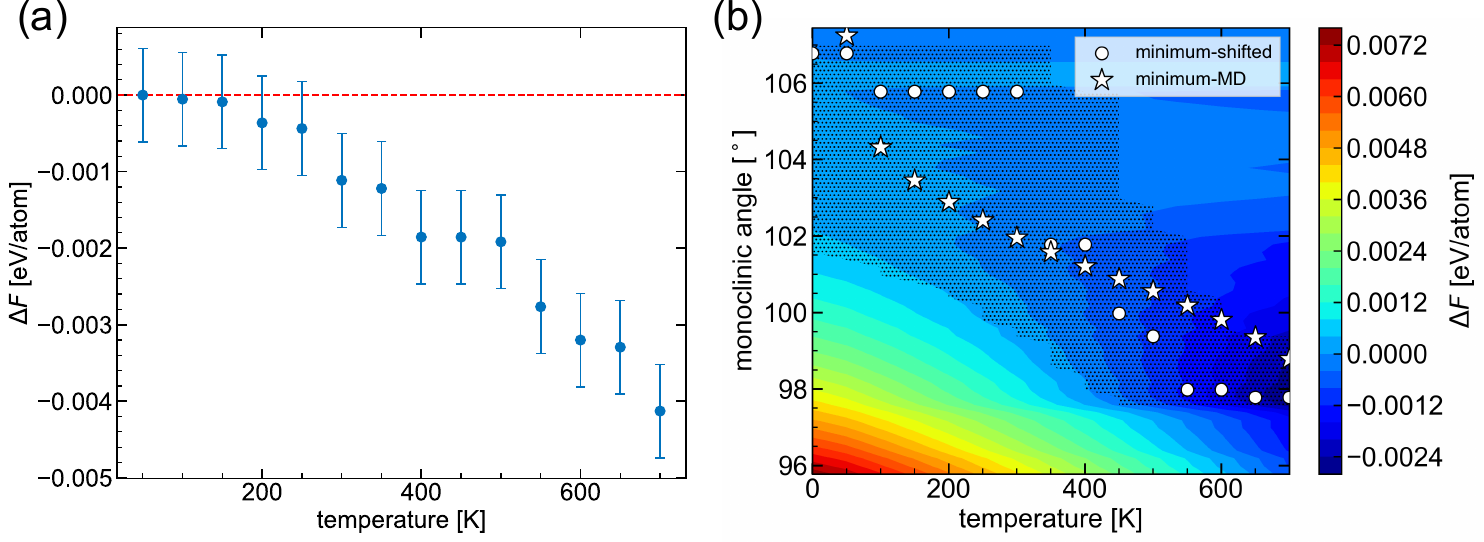}
    \caption{(a) Difference between the free energy of the mean structures obtained from HDNNP-based MD simulations and the free-energy minima derived from monoclinic shifts at temperatures ranging from 50 to 700~K. The negative values indicates that the MD-derived mean structures summarized in \autoref{tab:Table_4} are consistently more energetically favorable. (b) Free-energy surface as a function of temperature and monoclinic angle for the shifted structures. The free-energy minima at each temperature are marked by white circles. For comparison, the temperature dependence of the monoclinic angle obtained from MD-derived mean structures is shown by asterisks.}
    \label{fig:figure_14}
\end{figure}

Although the previous analysis of the shifted structures was able to predict the experimentally measured monoclinic angle, MD simulations performed with the same HDNNP potential require an approximately 300~K higher temperature to stabilize a monoclinic angle of 98.7$^\circ$. A similar discrepancy in the relationship between monoclinic angle and temperature has also been reported in DFT-based studies \cite{haskins_ab_2016,haskins_finite_2017}. Since MD at a given temperature represents an unconstrained search for structures with minimal free energy, the mean structures obtained from MD and summarized in \autoref{tab:Table_4} are expected to have lower free energies at a given temperature than those corresponding to the minima on the connecting path composed of shifted structures shown in \autoref{fig:figure_13}. \autoref{fig:figure_14} compares the free-energy minima obtained at each temperature along the connecting path in \autoref{fig:figure_13} with those obtained from the MD simulations. The resulting differences \autoref{fig:figure_14}a indicate that, at a given temperature, the MD-relaxed structures are thermodynamically more favorable.

This model builds on previous DFT studies reported in \cite{haskins_ab_2016, haskins_finite_2017, ko_temperature_2018}, which investigated the role of vibrational entropy in stabilizing the B19$'$ phase. The HDNNP framework overcomes key limitations of conventional DFT by enabling simulations of larger systems over longer timescales. Unlike reaction-coordinate-based approaches, HDNNP-driven molecular dynamics simulations allow free-energy minima to be identified while simultaneously accounting for all atomic degrees of freedom and thermal expansion effects within a single simulation framework.

The developed HDNNP model suggests that the B33 phase undergoes a temperature-driven instability, manifested by a gradual evolution of the monoclinic angle through a shear-like mechanism. The present study focuses primarily on the ideal equiatomic composition, excluding defects, external pressure, and applied stress. Nevertheless, in both experimental and theoretical studies, the monoclinic angle is known to be highly sensitive to deviations from stoichiometry, vacancies and antisite defects \cite{li_effects_2024, mizuno_compositional_2015, frenzel_effect_2015}, as well as to structural interfaces such as twins \cite{sestak_effect_2014}. It is also affected by hydrostatic pressure due to the associated volume changes \cite{bakhtiari_ab_2020, wang_resolving_2012}, and by applied shear stresses as a consequence of the low shear elastic constant $C_{55}$ of the martensitic phase \cite{wagner_lattice_2008, huang_crystal_2003}.

\begin{figure}[h!]
    \centering
    \includegraphics[width=1\textwidth]{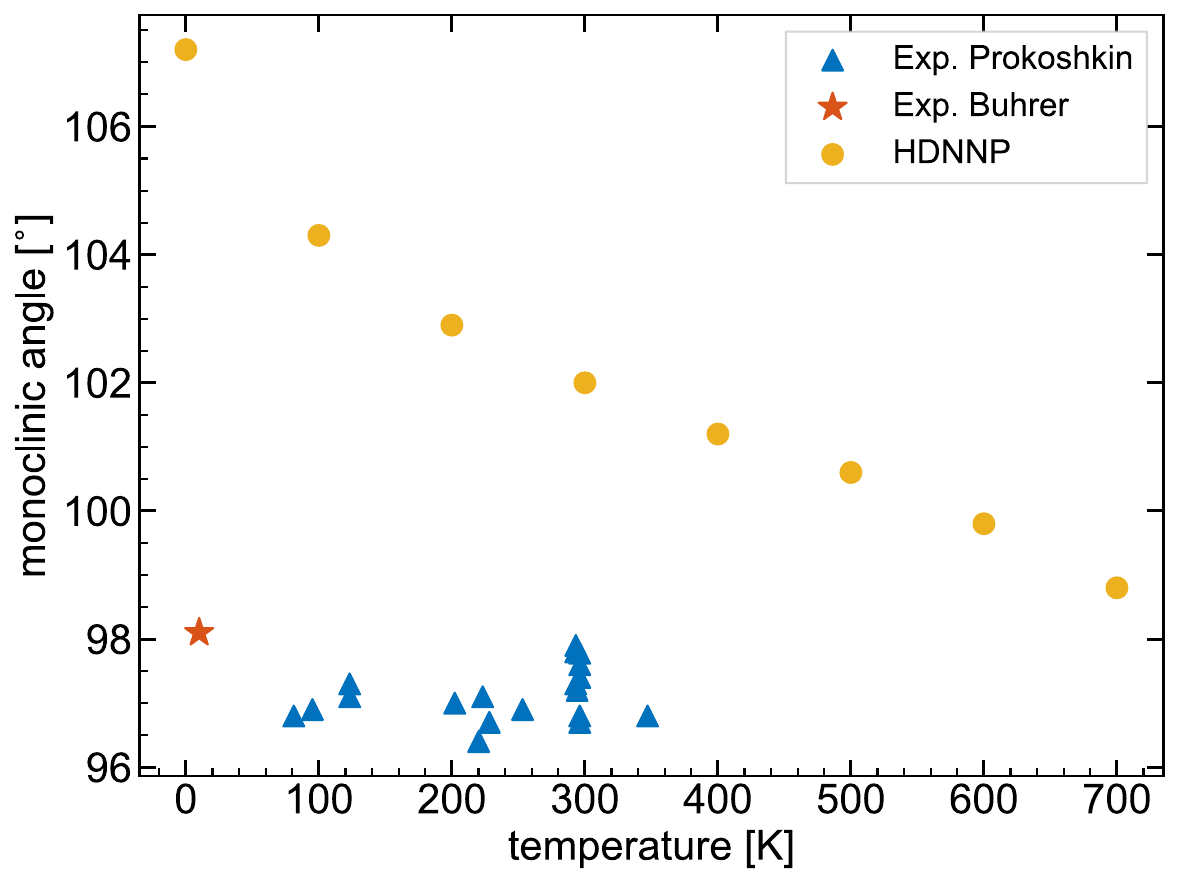}
    \caption{Comparison of the temperature evolution of the monoclinic angle predicted by HDNNP with experimental data taken from Prokoshkin \cite{prokoshkin_lattice_2004} and Buhrer \cite{buhrer_powder_1983}.}
    \label{fig:figure_15}
\end{figure}

Although both DFT and HDNNP calculations consistently indicate a sensitivity of the monoclinic angle to temperature \cite{haskins_finite_2017, haskins_ab_2016}, the experimentally observed monoclinic angle appears comparatively stable over the investigated temperature range, as shown in \autoref{fig:figure_15}. This discrepancy suggests that additional compositional, structural, or mechanical factors may contribute to the experimentally measured response. These effects will be systematically addressed in future work by extending the model to include representative atomic configurations that explicitly capture deviations from stoichiometry, structural interfaces and defects.

\section{Conclusion}
This paper presents a new interatomic potential suitable for simulations of the  B19$'$  martensitic structure in NiTi shape memory alloy. The potential is formulated in the framework of High Dimensional Neural Networks. The potential is based on an extensive DFT dataset that maps the low-energy martensitic atomic configurations in detail. The accuracy of the potential relative to the reference DFT data is better than 1 meV/atom. This accuracy is necessary to capture the subtle energy changes associated with the stability of the B19$'$ structure at finite temperatures relative to the orthorhombic B33 structure and to simulate structural evolution with temperature.

Several properties of the martensitic structure obtained using HDNNP were compared with direct DFT calculations. This comparison provides insight into the quality of the fit of the entire energy landscape and the suitability of the potential to capture lattice dynamics effects. The potential accurately reconstructs the ground state structure of B33, the energy evolution along the B33 to  B19$'$  path, the generalized stacking fault energy surface, the elastic properties of both B33 and  B19$'$  phases, and the phonon dispersion. This comparison indicates that the potential reconstructs relevant properties for simulations of finite-temperature phenomena with DFT accuracy and that it can be considered as a method for upscaling DFT simulations.

The presented dynamic simulations exceed the current limits of dynamic simulations of the  B19$'$  structure with DFT accuracy. Although they show the thermal evolution of the martensitic structure in agreement with previous DFT calculations. There is an important difference between them: The stability of the  B19$'$  structure at finite temperatures has so far been rationalized using DFT-based free energy calculations along a simply parameterized connecting path from B33 to  B19$'$, described by the monoclinic angle, or by ab-initio MD on very small atomic systems. Here we present simulations of a system containing 32,000 atoms over a few tens of ps timescales, where no prior assumption of its structural evolution was made. The simulations provide a theoretical prediction of the entire structural evolution of  B19$'$  martensite with temperature, including not only the evolution of the monoclinic unit cell angle, but also the evolution of lattice vectors and atomic positions.

The construction of this potential is an important first step for further atomistic simulations of the exceptional properties of the  B19$'$  structure, such as the presence of the easy [100]$_M$(001)$_M$ slip, which is unusual for intermetallic martensitic systems. The presented HDNNP potential for the B19$'$ martensitic phase is available at \cite{model_zenodo} in a form compatible with the LAMMPS and n2p2 packages.

\section*{Funding}

This work has been financially supported by the Czech Science Foundation [project No. 25-16285S] and by the
Operational Programme Johannes Amos Comenius of the Ministry of Education, Youth and Sport of the Czech Republic, within the frame of project Ferroic Multifunctionalities (FerrMion) [project No. CZ.02.01.01/00/22\_008/0004591], co-funded by the European Union. We also acknowledge the Ministry of Education, VSB – Technical University of Ostrava, IT4Innovations National Supercomputing Center, Czech Republic supported by the Ministry of Education, Youth and Sports of the Czech Republic through the e-INFRA CZ (ID:90254). This work was financially supported by the Student Grant Competition of the Czech Technical University in Prague (SGS), Project No. SGS24/146/OHK4/3T/14.

\section*{CRediT author contributions}
\textbf{Petr Jaroš:} Conceptualization, Methodology, Investigation, Data curation, Visualization, Writing – original draft.
\textbf{Petr Sedlák:} Conceptualization, Methodology, Supervision,, Data curation, Writing – review and editing.
\textbf{Petr Šesták:} Methodology, Validation, Formal analysis, Writing – review and editing.
\textbf{Miroslav Černý:} Methodology, Validation, Writing – review and editing.
\textbf{Jörg Behler:} Methodology, Software, Supervision, Writing – review and editing.
\textbf{Hanuš Seiner:} Conceptualization, Supervision, Writing – review and editing.

\section*{Declaration of competing interest}
The authors declare that they have no known competing financial
interests or personal relationships that could have appeared to influence
the work reported in this paper.

\section*{Data availability}

The high-dimensional neural network potential developed in this work, including the files required for its use with LAMMPS and n2p2, is publicly available through Zenodo at \url{https://doi.org/10.5281/zenodo.20342518}. Additional data supporting the findings of this study are available from the corresponding author upon reasonable request.

\appendix
\section{High-Dimensional Neural Network Potentials}
\label{appendix_1}
The HDNNP  formalism \cite{behler_generalized_2007, behler_4thgeneration} is based on a decomposition of the total energy of the system into contributions associated with the local atomic environments. Local atomic environments are described using atomic-centred symmetry functions (ACSFs)~\cite{behler_atom-centered_2011}, which characterize the neighborhood of each atom within a finite cutoff radius $R_c$. The cutoff function $f_{\mathrm{c}}^i(R_{ij})$ for a central atom $i$ is defined as
\begin{equation}
\label{equation_5}
f^i_{\mathrm{c}}(R_{ij}) =
\begin{cases}
0.5\,[\cos(\pi x)+1] &  R_{ij} \leq R_c \\
0 & R_{ij} > R_c
\end{cases}
\end{equation}
where $ x = R_{ij}/{R_c } $ and $R_{ij}$ denotes the distance between atoms $i$ and $j$. The local environments are encoded using a combination of radial and angular symmetry functions. The radial symmetry functions are given by
\begin{equation}
\label{equation_6}
G^2_i = \sum_{j} e^{-\eta R_{ij}^2} f_{\mathrm{c}}(R_{ij}),
\end{equation}
while the angular symmetry functions are defined as
\begin{align}
\label{equation_7}
G^{3}_{i} &=
2^{1-\zeta}
\sum_{j}\sum_{k}
\Big[
(1+\lambda \cos\theta_{ijk})^{\zeta}
e^{-\eta (R_{ij}^2 + R_{ik}^2 + R_{jk}^2)}
f_{\mathrm{c}}(R_{ij})
f_{\mathrm{c}}(R_{ik})
f_{\mathrm{c}}(R_{jk})
\Big],
\end{align}
where $\eta$, $\zeta$, and $\lambda$ are tunable parameters controlling the spatial resolution and angular sensitivity of the descriptors, and $\theta_{ijk}$ denotes the angle formed by atoms $i$, $j$, and $k$. These parameters were selected to ensure a balanced and uniform sampling of the local atomic environments within the cutoff radius, following the guidelines proposed in~\cite{behler_atom-centered_2011}. The resulting descriptor values are used as inputs to a fully connected feedforward neural network that predicts atomic energy contributions. The total energy of the system is obtained as the sum over atomic energy contributions, while the acting force is calculated as a derivative of the neural network defined  as
\begin{equation}
\label{equation_8}
F_{\alpha} = -\frac{\partial E}{\partial \alpha} = -\sum_{j=1}^{N_{\text{atom}}} \frac{\partial E_j}{\partial \alpha} = -\sum_{j=1}^{N_{\text{atom}}}\sum_{\mu=1}^{N_{\text{sym},j}} \frac{\partial E_j}{\partial G_{j\mu}} \frac{\partial G_{j\mu}}{\partial \alpha}
\end{equation}
where $\alpha$ denotes to Cartesian coordinate, $N_\text{atom}$ represents the total number of atoms, $N_\text{sym}$ goes through ACSFs $G_{j\mu}$. The fitting quality is measured by computing the root-mean-square errors (RMSE) for energies and forces component according to
\begin{equation}
\label{equation_9}
    \text{RMSE}(E) = \sqrt{\frac{1}{N_{\mathrm{struct}}}\sum_{i=1}^{N_{\mathrm{struct}}}(E_{\mathrm{NN}}^i - E_{\mathrm{ref}}^i)^2}
\end{equation}
\begin{equation}
\label{equation_10}
    \text{RMSE}(F) = \sqrt{\frac{1}{N_{\mathrm{struct}}}\sum_{i=1}^{N_{\mathrm{struct}}}\sum_{j=1}^{3N_{\mathrm{atom}}^i} (F_{j,\mathrm{NN}}^i - F_{j,\mathrm{ref}}^i)^2}
\end{equation}

where \(N_{\mathrm{struct}}\) denotes the number of structures included in the dataset, the subscript \textit{NN} represents the neural network predictions, and subscript ref represents values obtained from DFT calculations.
\clearpage
\FloatBarrier
\printbibliography

\end{document}